# Developing a Meta-suggestion Engine for Search Queries

**Seungmin Kim[1], EunChan Na[1], Seong Baeg Kim[1]**
[1]Department of Computer Education, Jeju National University, Jeju 63243, Republic of Korea

Corresponding author: Seong Baeg Kim (sbkim@jejunu.ac.kr)

This research was supported by the MSIT(Ministry of Science and ICT), Korea, under the National Program for Excellence in SW (2018-0-01863), supervised by the IITP(Institute for Information & communications Technology Promotion).

**ABSTRACT** Typically, search engines provide query suggestions to assist users in the search process. Query suggestions are very important for improving users' search experience. However, most query suggestions are based on the user's search logs, and they can be influenced by infrequently searched queries. Depending on the user's query, query suggestions can be ineffective in global search engines but effective in a domestic search engine. Conversely, it can be effective in global engines and weak in domestic engines. In addition, log-based query suggestions require many search logs, which makes them difficult to construct outside of a large search engine. Some search engines do not provide query suggestions, making searches difficult for users. These query suggestion vulnerabilities degrade the user's search experience. In this study, we develop a meta-suggestion, a new query suggestion scheme. Similar to meta-searches, meta-suggestions retrieves candidate queries of suggestions from other search engines. Meta-suggestions generate suggestions by reranking the aggregated candidate queries. We develop a meta-suggestion engine (MSE) browser extension that generates meta-suggestions. It can provide query suggestions for any webpage and does not require a search log. Comparing our meta-suggestions to major search engines such as Google, showed a 17% performance improvement on normalized discounted cumulative gain (NDCG) and a 31% improvement on precision. If more detailed factors, such as user preferences are discovered through continued research, it is expected that user searches will greatly improve. An enhanced user search experience is possible if factors, such as user preference, are examined in future work.

## I. INTRODUCTION

During internet information retrievals, web users search multiple times to obtain their desired data during the retrieval process. The user repeatedly expands their query and search until they obtain the data they want. Building queries that contain all a user's requirements is difficult [1]. To help users expand queries, search engines typically provide query suggestion services. Query suggestion is a representative search-related technology that provides related queries so that search users can build effective queries in the retrieval process [2], [3]. Query suggestion is effective in improving the user's search experience.

However, not all search engines used by users support query suggestions. Query suggestions typically use a user's query log to generate candidate queries [4]-[6]. Log-based query suggestions have a limitation; they can be contaminated by queries that are not included in the log.

Additionally, some search engines do not support query suggestions at all. Due to differences in the suggestions of various search engines, it is difficult for users searching on multiple search engines to find data.

To solve these problems, in this paper we propose meta-suggestions, a new type of query suggestion that does not require a query log. A meta-suggestion applies the ideas of meta-search to query suggestions. Similar to meta-search, which retrieves the initial query from the target engine and obtains the search results, a meta-suggestion obtains candidate queries from the target engine [7]. Meta-suggestions are useful for search engines where it is difficult to collect user data logs and create models. We develop the meta-suggestion engine as a browser extension. We divided the meta-suggestion process into two main steps consisting of suggestion generation and proposal suggestion reranking [8].

In the first step, suggestion generation, we retrieve candidate queries through a meta-search. A search on predefined target engines and suggested queries in the target engines is used as a candidate query. Next, the suggestion reranking reranks the candidate queries and displays the top-ranked queries to the user. Query reranking is essential, as the number of displayable queries is limited. We rerank the queries into an algorithm based on the evaluation factors we define.

We develop a meta-suggestion engine (MSE) based on the meta-suggestion. MSE is a browser extension that works on any webpage. We evaluate the developed MSE with our own dataset. It is evaluated using a new metric, the average hit rank (AHR), based on the prediction hit rate, and a classic metric, the normalized discounted cumulative gain (NDCG) [9]. We compare our engine to the major search engines. MSE shows a performance improvement of approximately 31% on precision and approximately 17% on NDCG.

Our study's research contributions are as follows. First, we propose a simple and effective new query suggestion scheme through meta-suggestions and query reranking. In addition, we develop MSE as a browser extension that can provide query suggestions wherever users search. Finally, we compare the proposed meta-suggestion with query suggestions of the major search engines and report a very good performance.

The remainder of this paper is structured as follows: Section 2 outlines the previous work on query suggestions and query expansions, and presents a literature search. Section 3 defines the evaluation elements and tasks used in the crawling process and the reranking algorithm. Section 4 describes the meta-suggestion design and implementation. Section 5 presents the experimental results and the conclusions, and issues related to future work are summarized in Section 6.

## II. RELATED WORK

### A. USER QUERY LOG BASED QUERY SUGGESTION

User query log-based query suggestions are one of the main subfields of query suggestions [4], [10]-[12]. User log-based query suggestions learn the user's query log, and then generate relevant queries as suggestions. [13] provides personalized query suggestions that reflect long-query and short-query actions based on a user's browsing history. In addition, there is a query suggestion method for modeling by additionally utilizing the clicked document log [14]. Approaches using demographic data or RRN-based approaches have also been proposed [15]. In query log-based query suggestions, collecting user data logs is essential. However, collecting and accessing the log is difficult due to the privacy and revenue concerns of search engines [16]. Therefore, user query log-based query suggestions are not feasible except for use in large search engines capable of collecting data. They are also malleable by queries not included in the log.

### B. QUERY SUGGESTION WITHOUT USER QUERY LOG

Recently, several studies have studied query suggestions without user query data. [16] proposed a study to anonymize data logs on a per-session basis to address concerns about user privacy that hinder data sharing. [17] developed a method for generating synthetic logs from publicly available datasets and utilizing them for evaluation. Many publications implement data other than user query data for model embedding. [8] addressed a method of generating suggestions by extracting phrases from a corpus as a method of removing the ambiguity of web queries with various interpretations. [18] proposed a query suggestion module that extracts queries using freely accessible online content, such as Wikipedia. Currently, query suggestions without a user query log provide query suggestions by training a model on a public dataset. However, query suggestions using corpuses, synthetic logs, etc. are malleable by queries not included in the data, in the same way as those based on user logs. However, there is a possible variation from these techniques in that we do not generate candidate queries from a predictive model but aggregate queries from other search engines.

### C. META-SEARCH

Meta-search is a search technique that implements an efficient search process. The key idea of a meta-search is to search for keywords in multiple search engines at the same time and to sort and return the generated results. We apply the ideas from meta-searches to query suggestions. The work of a traditional meta-search focuses on the selection of a sorting strategy and target engine for the retrieved search results [19], [20]. [21] proposed a target engine selection method based on a combination of a weighted round-robin algorithm and an artificial neural network. [22] reranked the result documents with the lion algorithm based on the fuzzy integrated extended nearest neighbor (FENN) classifier. To our knowledge, there are no studies using meta-search for query suggestions. In this paper, we propose a meta-suggestion-based query suggestion that obtains candidate queries from multiple target engines and returns the reranked queries.

## III. META-SUGGESTION

When a user searches using a search engine, most search engines typically generate query suggestions and display them to the user. Not all search engines support query suggestions. Query suggestions based on user query data can be influenced by queries that are not included in the query log. For example, the Korean search engine Naver does not provide query suggestions for most English search terms. In addition, Yahoo, which is not very popular in Korea, does not work well with Korean queries.

In this study, we define a new query suggestion scheme, meta-suggestion. A meta-suggestion is a new query suggestion scheme that aggregates query suggestions from other search engines as candidate queries for the suggestions. Meta-suggestions are effective for various initial queries because they fetch candidate queries from multiple target engines. For example, if a user searches with an English query, the meta-suggestion aggregates candidate queries from an English search engine such as Google and provides them to the user. Conversely, if a user searches with a Korean query, Meta-suggestion can aggregate candidate queries from Naver, a Korean search engine.

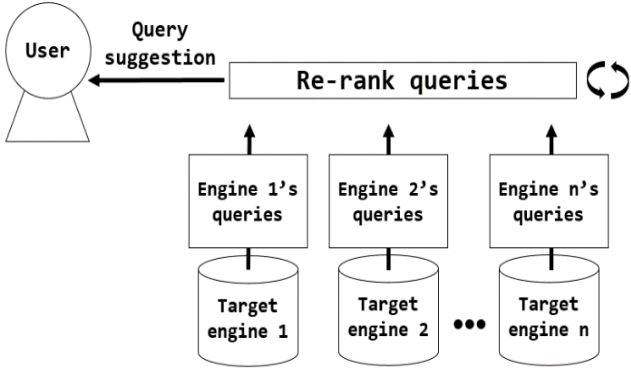

**FIGURE 1.** The architecture of Meta-suggestion

Figure 1 shows the architecture for the meta-suggestion. When a user searches with an initial query, the meta-suggestion takes the query suggested by the target engine for the initial query as a candidate query. Next, it reranks the candidate queries and returns the top queries. Finally, query suggestions show the user the queries up to the top cutoff.

### A. QUERY GENERATION
When a user searches, the meta-suggestion takes as input an initial query, which is a query typed by the user. Meta-suggestions request query suggestions from predetermined target search engines. The list of default target engines contains the top 5 Korean search engines, Naver, Google, Daum, Bing and Yahoo. The aggregated candidate queries are preprocessed. Duplicates are removed between candidate queries, and the count of duplicates as property of the query are calculated. Then, the query properties are used for query reranking. The preprocessed candidate query has the same properties as listed in Table 1.

### B. SUGGESTION RE-RANKING
It is not helpful to display all the candidate queries because the number of candidate queries are numerous at approximately 30 or more. If we display all queries to a user, the precision of the query suggestions will be increased. However, it reduces the recall ratio of the query suggestions and degrades the readability of the service. Therefore, it is

**TABLE 1.** Candidate suggestion attributes

| Property name | Data type | Description |
|---|---|---|
| **Name** | String | Candidate query word |
| **Loc** | String | Name of target engine |
| **Rank** | Int | Ranking of search queries on target engine |
| **NoD** | Int | Total number of duplicates between candidate queries |
| **Similarity** | Float | Similarity to the initial query |

necessary to rerank all fetched candidate queries and just show the top-ranked queries. The evaluation factors considered in the MSE design to rerank the candidate queries are as follows:

PRIORITY 1) DUPLICATION OF CANDIATE QUERY
In the meta-suggestion as a search aggregator that uses the data of several search engines to produce our own results, it is inevitable that duplication of the queries fetched from the search engines will occur. However, this duplication can also be evidence of the importance of the query. MSE uses duplicates of queries as evaluation factors.

First, we define $SUG_e(q)$ as engine $e$'s suggestion list of the initial query $q$. $SUG_e(q)$ appears as a set of candidate queries $q'$.

$$SUG_e(q) = \{ q'_1, q'_2, q'_3 \dots q'_{cutoff} \} \quad (1)$$

We obtain a number of duplicates $D(q')$, for all the candidate queries. $D(q')$ ranges from 1 to the number of target engines. If the set of target engines is $E$, the duplication $D(q')$ of candidate query $q'$ is calculated as follows:

$$IsHave(e, q') = \begin{cases} 1, & \textit{if } q' \textit{ in } SUG_e(q) \\ 0, & \textit{otherwise} \end{cases} \quad (2)$$

$$D(q') = \sum_{e \in E} IsHave(e, q') \quad (3)$$

PRIORITY 2) QUERY RANK IN TARGET ENGINE
In most cases of a query suggestions, a candidate query with a high score is at the top of the list, which allows the users to easily access them, and a query with a low score is at the bottom. Therefore, we consider the candidate query ranking in the target engine as an evaluation factor of the meta-suggestion. A lower rank value is better, and the rank ranges from 0 to less than the number of candidate lists by the engine.

Assuming that the number of target engines is $n$ and the rank of candidate query $q'$ in target engine $e$ is $R(e,q')$, the query rank $R(q')$ is calculated as follows:

$$R(q') = min(R(q', e_1), \dots R(q', e_{n-1}), R(q', e_n)) \quad (4)$$

PRIORITY 3) SIMILARITY WITH THE SEARCH QUERY

Previous research has shown that the user's next query in the search session of the user search process is similar to the user's previous query. Therefore, we select a query similar to the initial query as the query suggestion.

Assuming that the length of the initial query $q$ is $L(q)$, the length of the candidate search query $q'$ is $L(q')$, and the number of identical characters between the $q$ and $q'$ is $C(q, q')$. The similarity $S(q')$, of the $q'$ is calculated as follows:

$$S(q') = \frac{C(q,q')}{max(L(q),L(q'))} \times 100 \qquad (5)$$

The candidate queries aggregated from several target engines will be refined using three factors. They will be applied in a sequence by priority, Factor 1, Factor 2 and then Factor 3. That is, Factor 1 is first applied to the candidate queries, and then we obtain queries with the same rank. As the second step, Factor 2 is applied to the intermediate results obtained by applying Factor 1. Then, we still obtain the results with the same ranks. As the third step, Factor 3 is applied to the results obtained from Factor 2. Finally, the candidate queries are sorted.

It then returns query suggestions up to a preset cutoff rank and displays them to the user. Determining the optimal cutoff value is a major challenge for query suggestions. If the cutoff value is too high, the query suggestion is less likely to provide the desired query. However, if the cutoff value is too low, the query suggestion will be less accurate. As the cutoff value increases, the probability that the user will provide the desired query, i.e., reproducibility, increases. We decided to set the cutoff value to 8 based on our evaluation results from the experiments described below.

Algorithm 1 is the pseudocode of the meta-suggestion algorithm (MSA).

## IV. DESIGN and IMPLEMENTATION

### A. DESIGN

The presence or absence of query suggestions has a significant impact on users' search efficiency. Suppose the user wants to buy "earphones". He searches on e-commerce search engines such as Amazon to purchase the product. Amazon does not provide query suggestions, so he has trouble expanding their queries. However, if Amazon can suggest queries such as "earphones jbl" and "earphones apple", it can effectively speed up the user's search process.

In this paper, we develop a meta-suggestion engine (MSE) through a web browser extension program to provide query suggestion services. It is difficult to collect users' search logs and generate the candidate query in a web browser extension. We solve this problem by applying meta-suggestions. A meta-suggestion provides query suggestions on any web page and requires no user query data.

**Algorithm 1:** Meta-suggestion Algorithm (MSA)

**Input:** QUERY
**Output:** CANDIDATE_QUERY

1: **when user search**:
2:     cutoff = set_cutoff;
3:     engines= [selected_search_engines]
       //Google, Naver, Daum, etc..
4:     result_queries= []
5:     **for** engine in engines
6:         c_queries = get_c_query(engine)
           //get candidate query from engine
7:         compare_similarity(query, c_queries)
8:         result_query.append(c_queries)
9:         check_duplicate(result_query)
10:    result_query.sort()
11:    **return** result_query[:cutoff]

The basic features of the MSE are:
- ✓ It maintains the query suggestion of the user's current search engine and displays additional query suggestions.
- ✓ If there is a duplicate query suggestion on the current web page and MSE, the duplicated suggestion is highlighted.
- ✓ Up to 8 queries are displayed.
- ✓ It provides the ability for users to turn on/off the entire function of the extension or turn on/off partial functions of the extension.
- ✓ A user can add or change the target engine of the MSE.

The entire MSE process ranges from starting a web browser to displaying the query suggestion. It consists of the following steps:

1. Start a web browser with the meta-suggestion extension program.
2. A user tries to determine the information that he or she wants.
3. The search query typed by the user and the searched URL are taken as input values.
4. Create a target engine URL to be crawled with the retrieved keywords and crawl the suggested keywords. In this process, the crawling activity for each target engine is performed simultaneously using asynchronous processing.
5. Apply the proposed meta-suggestion algorithm.
6. If the suggestion on the current web page overlaps with a suggestion from MSE, highlight that query.
7. The returned suggested queries are displayed on the web page.

Figure 2 shows the operational process of the MSE according to the two situations. When a user searches, the web browser passes the initial query the user retrieves to MSE. MSE performs MSA with the received query as input and obtains a primary query suggestion. If the search engine the user is currently located in does not have query suggestions, it is displayed as is, but if the search engine also has query suggestions, the query suggestions of MSE and the query suggestions of the current engine are compared, and duplicate queries are highlighted and not displayed.

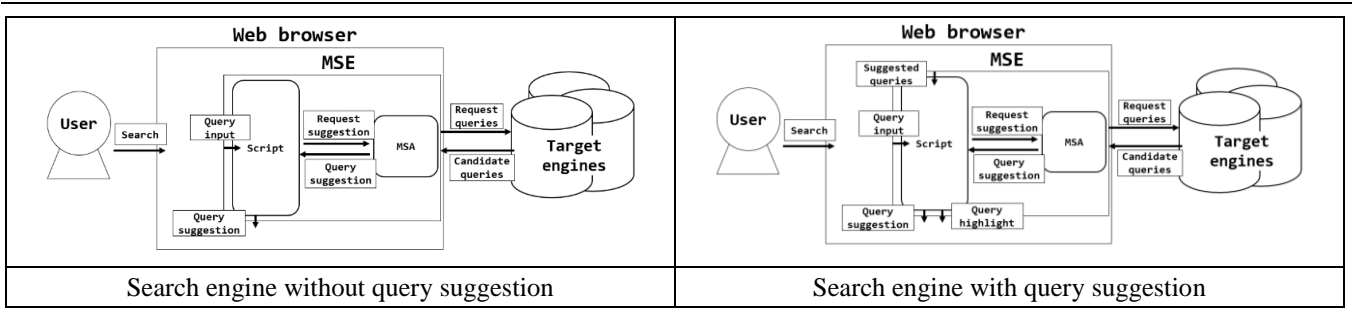

**FIGURE 2.** The abstraction of MSE for two situations

### B. IMPLEMENTATION

We implemented MSE as a browser extension. The contents and characteristics of the extension program are described in the order of operation of the extension program as follows:

If you install the extension first, the extension icon will appear on the right side of the search banner when you launch the Chrome web browser. Figure 3 at the bottom shows that the extension program icon appears when the extension program is installed.

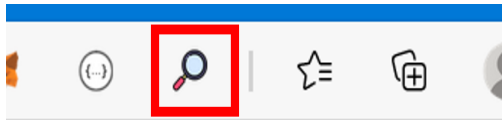

**FIGURE 3.** Extension icon snapshot

When you click the extension icon, the extension's pop-up page appears. On the pop-up page, the user can control the functions of the extension. You can start and end the search suggestion function or the query highlighting function through the toggle switch on the pop-up page. In addition, the user can add or delete target engines from which the extension will obtain candidate queries. Figure 4 and Figure 5 shows the GUI that controls the function of the extension program through the pop-up page.

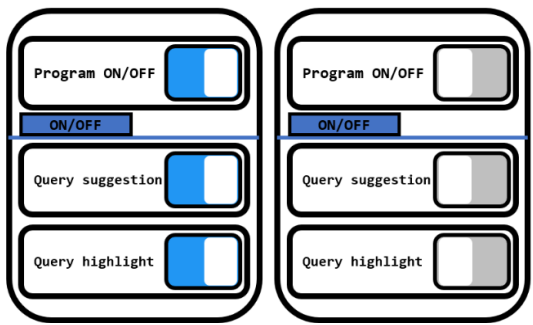

**FIGURE 4.** On/off extension interface

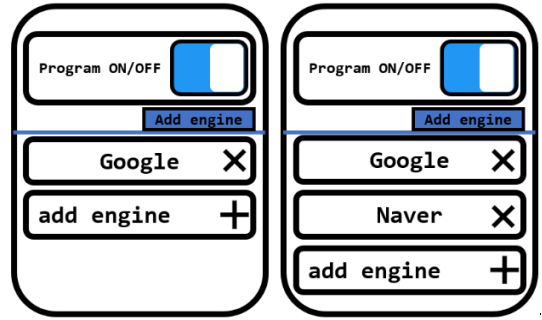

**FIGURE 5.** Target engine interface

After you set up query suggestions to run, MSE detects your searches. When a search is detected, the MSE runs the MSA and displays the resulting queries in the browser. If you click on the displayed query, you will be taken to the search results for the clicked query. Figure 6 is a snapshot of a web page where suggestions are displayed when a user searches for 'machine learning'.

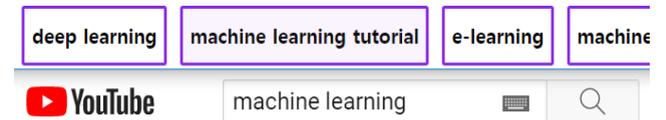

**FIGURE 6.** Extension program suggestion output snapshot

If you turn on the highlighting feature, MSE checks the results of the MSA and the duplicates of the suggestion on the current web page. If there are duplicates, it highlights the duplicate queries. Figure 7 is a snapshot with highlights when a user searches for 'Korea'.

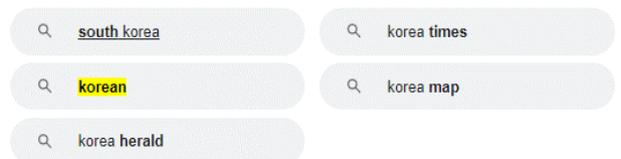

**FIGURE 7.** Query highlight function output screenshot

9

**TABLE 2.** Experimental results of MSE and baseline

| Engine | RECALL | PRECISION | AHR | NAHR | NDCG |
|---|---|---|---|---|---|
| **MSE@8** | 0.060 | 0.0075 | 2.53 | 0.316 | 0.820 |
| **MSE@10** | 0.069 | 0.0069 | 3.69 | 0.369 | 0.819 |
| **MSE@20** | 0.092 | 0.0046 | 6.50 | 0.325 | 0.832 |
| **Google** | 0.046 | 0.0057 | 2.75 | 0.343 | 0.700 |
| **Naver** | 0.008 | 0.0008 | 2.29 | 0.229 | 0.079 |
| **Daum** | 0.027 | 0.0013 | 3.66 | 0.183 | 0.565 |
| **Yahoo** | 0.010 | 0.0012 | 1.70 | 0.212 | 0.414 |
| **Bing** | 0.027 | 0.0033 | 3.04 | 0.38 | 0.658 |

## V. EVALUATION

### A. DATASET.

We adopt our own collected data as the experimental dataset. The dataset was collected from the search history of Korean users from March to September 2021. In total there were 5,604 queries collected. The search log includes timestamps, queries in English and Korean and user indices. In Table 2, we can see the amount of data we collected.

**TABLE 3.** The amount of dataset

| TOTAL NUMBER OF QUERIES | NUMBER OF KOREA QUERIES | NUMBER OF ENGLISH QUERIES | QUERY WITH BOTH |
|---|---|---|---|
| 5,604 | 3,437 | 1,287 | 880 |

We split our dataset into query sessions. A query session is a cluster of query logs starting from when the user initiates a search to when the user finds the data he/she needs. We used the DBSCAN algorithm to cluster the queries into sessions. DBSCAN is applied with *eps = 30, minPoints=1 and feature = timestamp*. We clustered 5604 queries with 1262 sessions as a result.

### B. EVALUATION METHODOLOGY

The goal of query suggestions is to help users write their next query. If the query suggestion predicts and displays the user's next query, users can shorten the query expansion and search faster. Therefore, we evaluate how well the suggestions can predict the next query that the user wants to enter [16].

We define query session $Qs$ as the set of queries searched by the user, as shown below:

$$Qs = \{q_1, q_2, q_3 \dots q_n\} \quad (6)$$

Additionally, we address the real next queries $After\_Qs(q_i)$ of the user, which is the comparison target of $SUG_e(q_i)$. $After\_Qs(q_i)$ is the set of queries placed after the user's initial query $q_i$ within the search session. The intersection between $SUG(q_i)$ and $After\_Qs(q_i)$ means that the engine's suggestion is predicting the user's next query.

$$After\_Qs(q_i) = \{q_{i+1}, q_{i+2} \dots q_n\} \quad (7)$$

$$Hit(q_i) = \begin{cases} 1 & if \ SUG(q_i) \cap After\_Qs(q_i) \\ 0 & otherwise \end{cases} \quad (8)$$

We use *the* Average Hit Rank (*AHR*), Normalized Average Hit Rank (*NAHR*), *Recall and Precision* as evaluation metrics. These metrics are calculated based on the *Hit* described above. *AHR* is the average of the ranks of queries $q_i$ when *Hit($q_i$) is 1*. AHR is the *AHR* normalized to the cutoff value. Assuming the entire query set is Q, *AHR* and *NAHR* are calculated as follows:

$$AHR = \frac{\sum_{q_i \in Q} \begin{cases} i & if \ Hit(q_i) \ is \ 1 \\ 0 & otherwise \end{cases}}{\sum_{q_i \in Q} Hit(q_i)} \quad (9)$$

$$NAHR = \frac{AHR}{cutoff} \quad (10)$$

*Recall* measures the percentage of query suggestions that are hit among all the query suggestions. *Precision* is *recall* divided by the cutoff value. When *no. Q* is the total number of queries, *recall* and *precision* are calculated as follows:

$$Recall = \frac{\sum_{q_i \in Q} Hit(q_i)}{no. Q} \quad (11)$$

$$Precsion = \frac{\sum_{q_i \in Q} Hit(q_i)}{no. Q \times cutoff} \quad (12)$$

Since these exact hits do not occur often and putting relevant queries at a top rank is important in query suggestions. We also use normalized discounted cumulative gain (NDCG), a relevance-based valuation metric. We measure relevance using Google's pretrained model BERT, which has shown promising results on several IR and natural language processing tasks [23].

### C. RESULT

We evaluated several versions of MSE@8, MSE@10 and MSE@20 with different *cutoff* values to compare differences according to the *cutoff* values. We compared MSE with major search engines in Korea. We considered Google, Naver, Daum, Yahoo and Bing as baselines. Additionally, we set them as target engines for MSE.

We evaluated the baseline and MSE with the previously described evaluation metrics. Table 2 shows the overall evaluation results. A high *recall* score indicates that the engine is better at predicting the user's next query. *Precision* shows results that reflect the number of query suggestions displayed by engines in *recall*. MSE@20 has the best *recall* score, but since the cutoff value is higher than MSA@8, the precision shows that MSA@8 has the best score.

*AHR* and *NAHR* are metrics of how high a search engine ranks the predicted query. In terms of *AHR*, Yahoo performs best, and Daum performs best in terms of *NAHR*. This is because most of the queries they hit were predictable queries.

The fact that engines with low recall values, such as Daum and ahoo, have high *AHR* values in common proves this. Given that MSE has the highest recall performance, the MSE's *AHR* and *NAHR* figures are reasonable.

In terms of NDCG, MSE also shows the best performance. Compared to Google, which has the best performance among engines, MSE shows a performance improvement of approximately 17 to 19 percent. Overall, we found that our method can generate higher-quality query suggestions than other search engines. Based on Google, the engine with the best performance index among the comparisons, MSE@8 showed a performance improvement of approximately 31% in *precision* and approximately, and 17% in NDCG.

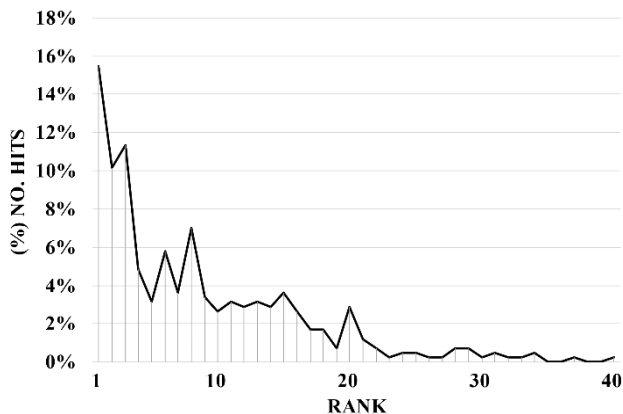

**FIGURE 8.** Ranking distribution of a query when the MSE succeeds in predicting the next query

Figure 8 shows the ranking distribution of a query when the MSE succeeds in predicting the next query. If the ranking algorithm is efficient, the graph appears as a downward-righting graph. MSE appears as a downward-sloping graph, which is in good agreement with expectations. The distribution of prediction hits has a huge impact on setting the cutoff value for query suggestions. An engine that shows as few offers as possible and hits as many predictions as possible is desirable. Looking at the distribution of MSE in Figure 8, 93% of the hits occur in candidate queries up to 20th. Next, up to 10th obtains 67% of the hits, and up to 8th obtains 61% of the hits. Considering the experimental results and that the major search engines Google and Bing use a cutoff value of 8, we decide to set the default cutoff value to 8.

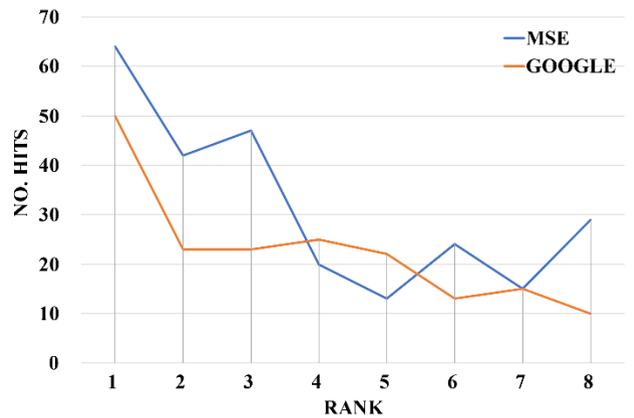

**FIGURE 9.** Distribution of successful predictions for the next query of MSE@8 and Google

Figure 9 shows the distribution of the successful predictions for the next query of MSE@8 and Google. Since Google showed the best performance among the baselines, it was set as a comparison target. As in Figure 8, if the ranking algorithm is efficient, the graph appears as a downward-righting graph. Both graphs appear as a downward-sloping curve but in the case of MSE, it increases further at 6 and 8. This shows that the ranking algorithm has room for further improvement compared to Google.

## VI. CONCLUSION

Traditional user log-based query suggestions cannot be used for small engines without logs and do not work well for queries not included in the log. In this study, we propose meta-suggestions, a new query extension scheme without user query logs. Meta-suggestions aggregate candidate queries from target engines similar to meta-search. We introduced three evaluation factors to evaluate the candidate queries brought into a meta-suggestion. We developed and evaluated the meta-suggestion engine, a browser extension that uses meta-suggestions. MSE can provide search term suggestions from any webpage, and no search logs are required. We evaluated a traditional relevance-based NDCG and a new actual predictive hit-based *precision*. Compared to Google's query suggestion, MSE showed a 17% performance improvement in NDCG and a 31% performance improvement in *precision*. However, our meta-suggestion has a limitation that performance may vary depending on the target engine. We will examine the selection of an effective target engine and improvement of the reranking algorithm in future studies.

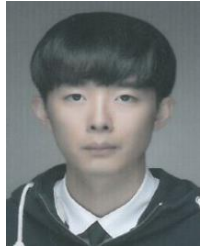

**SEUNGMIN KIM** pursued a B.S. degree at Jeju National University, Korea with a dual degree in Computer Education and Blockchain Security. His current research interests include blockchain, information security, and data science.

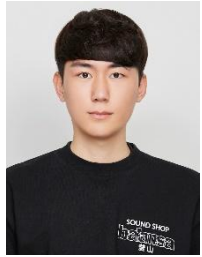

**EUNCHAN NA** is currently pursuing a B.S. degree in computer education at Jeju National University, Jeju, South Korea. His research interests include computer science education, educational games and computer architecture.

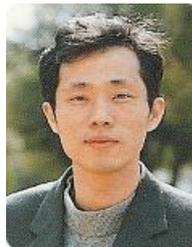

**SEONG BAEG KIM** received a B.S., M.S., and Ph.D. in Computer Engineering from Seoul National University, Korea, in 1989, 1991, and 1995, respectively. He is currently a professor at the Dept. of Computer Education at Jeju National University, where he has been since 1996. He was a visiting scholar at the Department of Computer Science, Montana State University from 2001 to 2002, Department of Electrical & Computer Engineering, University of Cincinnati from 2008 to 2009 and University of California, Berkeley from 2015 to 2016, respectively. His research interests include data science, computer systems, and computer science education.